\documentclass[preprint2]{aastex62}
\pdfoutput=1 
\usepackage{amsmath,amstext}
\usepackage[T1]{fontenc}
\usepackage{apjfonts} 
\usepackage{comment}
\usepackage[figure,figure*]{hypcap}
\usepackage{multirow}
\usepackage{amsmath}
\usepackage{nccmath}
\usepackage{psfrag,graphicx,color}
\usepackage{soul}

\setstcolor{red}

\shorttitle{VLA observation of PSR J1813-1749}
\shortauthors{Dzib et al.}

\begin{document}

\title{The Enigmatic compact radio source coincident with the energetic X-ray pulsar PSR~J1813$-$1749 and HESS~J1813$-$178}


\correspondingauthor{Sergio A. Dzib}
\email{sdzib@mpifr-bonn.mpg.de}

\author[0000-0001-6010-6200]{Sergio A. Dzib}
\affiliation{Max-Planck-Institut f\"ur Radioastronomie, Auf dem H\"ugel 69,
 D-53121 Bonn, Germany}
 \author[0000-0003-2737-5681]{Luis F. Rodr\'{\i}guez}
\affiliation{Instituto de Radioastronom\'{\i}a y Astrof\'{\i}sica, Universidad Nacional Aut\'onoma de M\'exico, Morelia 58089, Mexico}
\author{Ramesh Karupussamy}
\affiliation{Max-Planck-Institut f\"ur Radioastronomie, Auf dem H\"ugel 69,
 D-53121 Bonn, Germany}
\author[0000-0002-5635-3345]{Laurent Loinard}
\affiliation{Instituto de Radioastronom\'{\i}a y Astrof\'{\i}sica, Universidad Nacional Aut\'onoma de M\'exico, Morelia 58089, Mexico}
\affiliation{ Instituto de  Astronom\'{\i}a, Universidad Nacional Aut\'onoma de M\'exico, Apartado Postal 70-264, CdMx C.P. 04510, Mexico}
\author{Sac-Nict\'e X. Medina}
\affiliation{Max-Planck-Institut f\"ur Radioastronomie, Auf dem H\"ugel 69,
 D-53121 Bonn, Germany}

\begin{abstract}
New VLA detections of the variable radio continuum source VLA J181335.1$-$174957, 
associated with the energetic X-ray pulsar PSR~J1813$-$1749 and the TeV source 
HESS J1813-178, are presented. The radio source has a right circular polarization 
of $\sim50\%$, and a negative spectral index of $-1.3\pm0.1$, which show that it is 
non-thermal. The radio pulses of the pulsar are not detected from additional 
Effelsberg observations at 1.4~GHz made within one week of a VLA detection. 
This result would appear to support the idea  that the continuum radio emission detected with the VLA does not trace 
the time-averaged emission pulses, as had previously been suggested. We discuss other possible origins 
for the radio source, such as a pulsar wind, magnetospheric emission, and a low-mass star companion. 
 However, observations made at higher frequencies by Camilo et al. (in preparation) show that the VLA source is in fact the time-averaged pulsed emission and that the detection of the pulses had not been achieved because this is the most scattered pulsar known.
\end{abstract}


\nopagebreak
\section{Introduction}\label{sec:intro}

PSR J1813-1749 (=CXOU J181335.1$-$174957)  is the second most energetic pulsar 
in the Milky Way \citep{halpern2012}. 
It was discovered  and identified as a young pulsar { with a spin period of 44.7 ms} by  \cite{GH2009} based on Chandra X-Ray observations. 
Later \cite{halpern2012} used  a few Chandra  and  XMM-Newton  observations  to  determine  the spin-down rate of PSR J1813$-$1749 to be $\dot{P}=1.265\times10^{-13}$,
corresponding to a spin-down luminosity of $\dot{E}= 5.6\times10^{37}$~erg~s$^{-1}$, thereby establishing it as an energetic young pulsar. These values 
are only below  those measured for the Crab pulsar  \citep[e.g.,][]{halpern2012}\footnote{\cite{halpern2012} mentioned PSR J2022+3842 as the second most energetic pulsar,
however a recent revision of its spin period  by \cite{arumu2014} show that it is 48 ms instead of 24 ms, thus the energy of PSR J2022+3842 is below the energy of PSR J1813$-$1749.}.

PSR J1813-1749 is associated with one of the brightest and most compact 
objects located by the HESS Galactic Plane Survey \citep{aharonian2005}, the
TeV source HESS J1813-178.  This HESS source has been associated with continuum 
high-energy emission from X-rays to gamma rays \citep{reimer2008,ubertini2005,abdo2009,albert2006}.
Within the TeV extent of this source lies G12.82-0.02,
a  young relatively compact shell-type  radio  supernova remnant (SNR)  with  a  diameter  of 
$\sim2'$ \citep{brogan2005}. Deep X-Ray observations revealed that a pulsar wind nebula (PWN) is embedded in the SNR, and is powered by PSR J1813$-$1749 \citep[][]{helfand2007,funk2007,GH2009}.
While from observations alone, it is not clear if the TeV emission comes from the SNR or the PWN, \cite{fang2010} predict that the high energy
emission is produced mostly in the PWN.
 
\cite{halpern2012} firmly established that PSR J1813-1749 is a young pulsar 
(characteristic age $\sim 5600$ yr) from their multi-year X-ray observations. On the other hand, 
several radio observations carried out in the past have not revealed any radio pulsations. The first  made at 1.4 GHz on 2005 September
8 with the Australia Telescope National Facility (ATNF) Parkes telescope  set  an  upper  limit  on the flux  density of the radio pulsar as $S_{1.4}<0.07$~mJy  \citep{helfand2007}. \cite{halpern2012} did a new search for pulsed emission
using the Green Bank Telescope (GBT) in 2009 May 25 and 2009 August 17, and placed upper  limits  to  the  period-averaged  flux  density  at  2.0 GHz of ${S_{2.0}<0.013}$~mJy and ${S_{2.0}<0.006}$~mJy, respectively. 
The common explanation for the lack of radio pulses from high energy pulsars, is that the radio emission beam is narrower than those at higher energies, and thus the radio pulses miss the sightline toward Earth \citep{bra1999,watters2011}. However, it is also known that pulsars may be intermittent emitters, and so observations could miss the pulses if they are scheduled during periods of inactivity. However intermittency is normally associated with pulsars of characteristic age, $\tau_c \approx$ 1~Myr or older \citep{kramer2006,llm+2012,crc+2012}.

In  an  unexpected  result,  \cite{dzib2010}  reported, by the first time, the detection of a compact 
($\leq0\rlap{.}''2$)  radio continuum source (named VLA J181335.1$-$174957) with a flux 
density of $180\pm20$~$\mu$Jy at 4.86 GHz, and located within $0\rlap{.}''2$ of the 
Chandra position of PSR J1813-1749. The observation was made on 2006 February 25. They estimated that the probability of the compact radio source being a background source is 
$5\times10^{-5}$. Thus there is a high chance that the compact radio 
source is related to the pulsar and might correspond to the integrated emission of the radio pulses.
More recently, however, \cite{dzib2010} did not detect the source during a multi-wavelength radio 
observation on 2009 March 24, and they were the first to suggest that VLA~J181335.1$-$174957 could be an intermittent radio pulsar. The alternative is that this low signal to noise detection is spurious.

We present new Karl G. Jansky Very Large Array (VLA) observations to show that the compact source reported by \cite{dzib2010} is active again. We use these new and archival observations to constrain the nature of the radio continuum source. In addition, we report on the results from pulsar periodicity searches carried out on the observations using the Effelsberg 100m Radio Telescope. These observations were separated only by a week from the VLA observations, which increase the chance of detecting radio pulsations, if any. 

\begin{figure*}[!th]
   \centering
\includegraphics[height=0.54\textwidth, trim=43 400 80 50, clip]{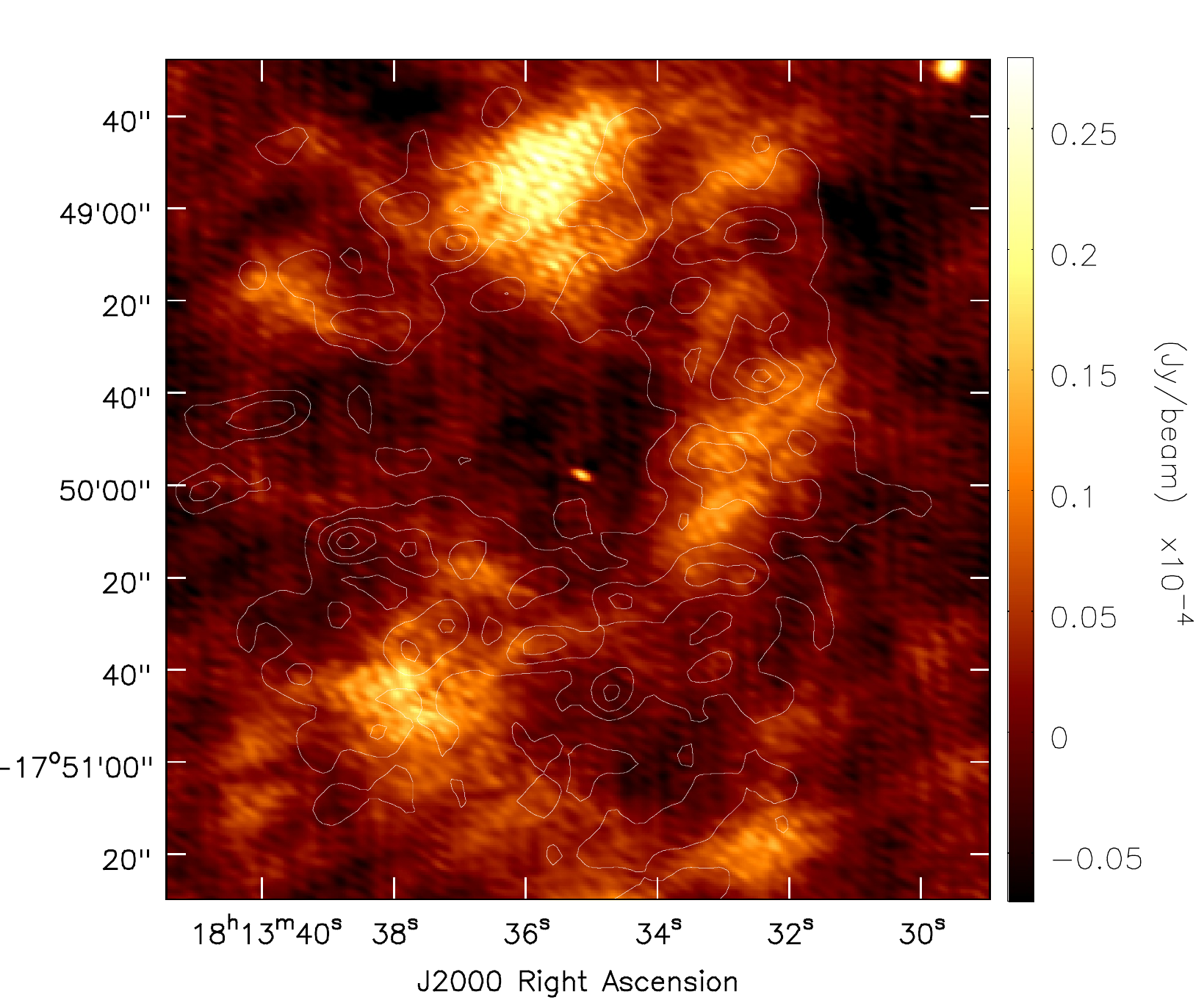}
   \caption{{\it Background:} C-band (5.5 GHz) radio image from 2012 May 6, with the compact source VLA J181335.1-174957 located
   at the center of the image. Diffuse emission from supernova remnant G12.82$-0.02$ is detected.
   {\it Contours:} 1.4~GHz radio emission of this SNR \citep[i.e.,][]{helfand2007}. This last 
   image was obtained from the \href{https://third.ucllnl.org/gps/index.html}{MAGPIS webpage} \citep{helfand2006}. }
   \label{fig:SNR}
\end{figure*}

\section{Observations and Data Calibration}

\subsection{VLA Observations}
In this paper, we use NRAO archive data of three unpublished observations in the direction of
PSR~J1813-1749. These observation were collected at 5.5 GHz as part of the VLA projects 12A-444 and 12B-278.
Additionally, we recently obtained new VLA observations at C-band (4--8 GHz) and X-band (8--12 GHz) during
project 17B-028. All the observations were recorded in standard wide-band continuum mode. The epochs,
and the observation details are listed in Table~\ref{tab:vlapsr}.

The data sets were edited, calibrated, and imaged using the Common Astronomical Software 
Applications package \citep[CASA;][]{mcmullin2007}. Specifically, we use the pipeline provided by 
the National Radio Astronomy Observatory (NRAO) to calibrate the data. The pipeline is  designed 
to flag strong Radio Frequency Interference (RFI) and to apply a standard calibration to the data. 
Images of the Stokes~I parameter  were performed using the task {\sl tclean}, with a weighting scheme 
intermediate between  natural and uniform. Similarly, we also imaged the Stokes V parameter
for most of the epochs, with the exceptions of the observations taken on 2012 October,
since the source was located far enough from the phase center that beam squint distortion (that 
introduces spurious circular polarization) is expected for this epoch. 

\begin{table*}
\small
\centering
\renewcommand{\arraystretch}{0.9}
\caption{Image results, and radio properties of VLA  J181335.1$-$174957.}
\begin{tabular}{lccccccccc}\hline\hline
Epoch       & $\nu$ & $\Delta\nu$& VLA&Phase & Beam  size              & Noise               &  $S_\nu$     &    &   $\pi_c$         \\
(yyyy.mm.dd)&(GHz)&(GHz)&Conf.&Calibrator&[$''\times'';\,\, ^{\circ}$]&($\mu$Jy bm$^{-1}$)&($\mu$Jy)&$\alpha$&(\%)\\
(1)&(2)&(3)&(4)&(5)&(6)&(7)&(8)&(9)&(10)\\
\hline
2012 May 06  & 5.5 & 2.0 & CnB& J1820-2528&$3.7\times1.8;$   64  & 2 &$20\pm3$  & ...& $50\pm13$     \\
2012 Oct. 04 & 5.5 & 2.0 &  A & J1733-1304&$0.49\times0.25;$ 16& 15&$74\pm15$ & ...      & ...  \\ 
2012 Oct. 06 & 5.5 & 2.0 &  A & J1733-1304&$0.52\times0.26;$ 25  & 13&$88\pm13$ & ...      & ...  \\ 
2017 Sept. 18& 6.0 & 4.0 &  B & J1811-2055&$1.51\times0.74;$ 15  & 8 &$112\pm9$ & \multirow{2}{*}{$-2.1\pm0.3$}& $44\pm6$  \\ 
2017 Sept. 18& 10.0& 4.0 &  B & J1811-2055&$0.95\times0.46;$ 18  & 6 &$38\pm6$  &  & $63\pm21$  \\ 
2017 Dec. 11 & 6.0 & 4.0 &  B & J1811-2055&$1.74\times0.77;$ 33  & 9 &$130\pm9$ & \multirow{2}{*}{$-0.9\pm0.2$} & $54\pm7$ \\ 
2017 Dec. 11 & 10.0& 4.0 &  B & J1811-2055&$1.18\times0.49;$ 35  & 7 &$83\pm7$  & & $28\pm8$   \\ 
2018 Jan. 8  & 6.0 & 4.0 &  B & J1811-2055&$1.49\times0.77;$ 23  & 7 &$124\pm7$ & \multirow{2}{*}{$-1.3\pm0.2$} & $45\pm6$ \\ 
2018 Jan. 8  & 10.0& 4.0 &  B & J1811-2055&$0.98\times0.47;$ 26  & 7 &$64\pm7$  &  & $41\pm10$ \\ 
2018 Jan. 13 & 6.0 & 4.0 &  B & J1811-2055&$1.76\times0.76;$ 32  & 8 &$123\pm8$ & \multirow{2}{*}{$-1.4\pm0.3$} & $47\pm7$ \\ 
2018 Jan. 13 & 10.0& 4.0 &  B & J1811-2055&$1.19\times0.47;$ 35  & 8 &$61\pm8$  &  & $51\pm15$ \\ 
2018 Jan. 21 & 10.0& 4.0 &  B & J1811-2055&$1.40\times0.51;$ 31  & 7 &$53\pm7$  &  ...& $59\pm13$ \\ 
2018 Jan. 28 & 6.0 & 4.0 &  B & J1811-2055&$1.44\times0.77;$ 16  & 7 &$131\pm7$ & \multirow{2}{*}{$-1.2\pm0.2$}  & $40\pm6$\\ 
2018 Jan. 28 & 10.0& 4.0 &  B & J1811-2055&$0.93\times0.48;$ 21  & 6 &$70\pm7$  &   & $46\pm10$\\ 
2018 Feb. 4  & 10.0& 4.0 & BnA& J1811-2055&$1.16\times0.25;$ -57 & 8 &$55\pm9$  & ...  & $49\pm18$ \\ 
\hline\hline
\label{tab:vlapsr}
\end{tabular}
\end{table*}

\newpage

\subsection{Effelsberg 100m Observations and pulsar search}
In addition to the VLA observations, we use the 100-m Effelsberg Telescope
to search for the radio pulses at 1.4~GHz. The source was observed on 2018 February 12 over 2$\times$24-minutes sessions with the central pixel of 7--Beam feed. The receiver was tuned to a frequency of 1.36~GHz and it provides an usable bandwidth of 240~MHz which are detected as two orthogonal polarizations. The coordinates for telescope pointing were taken from \cite{halpern2012}. The 240~MHz signal was recorded as 320~MHz baseband data using the PSRIX system \citep{lkg+2016} and then converted to a filter-bank format with 512-channels and a time resolution of 64~$\mu$s. After cleaning intermittent RFI, the above data was searched for any periodic signals with a standard pulsar search software, PRESTO \citep{ransom2001}. The search was sensitive to a dispersion measure (DM) of 3496, which is several times the DM estimated for a distance of 4.8~kpc by both NE2001 and YMN16 models for the Galactic electron-density \citep{ne2001,ymn2016}.

\section{Results}

\begin{figure}[!th]
   \centering
  \includegraphics[height=0.38\textwidth, trim=15 0 0 10, clip]{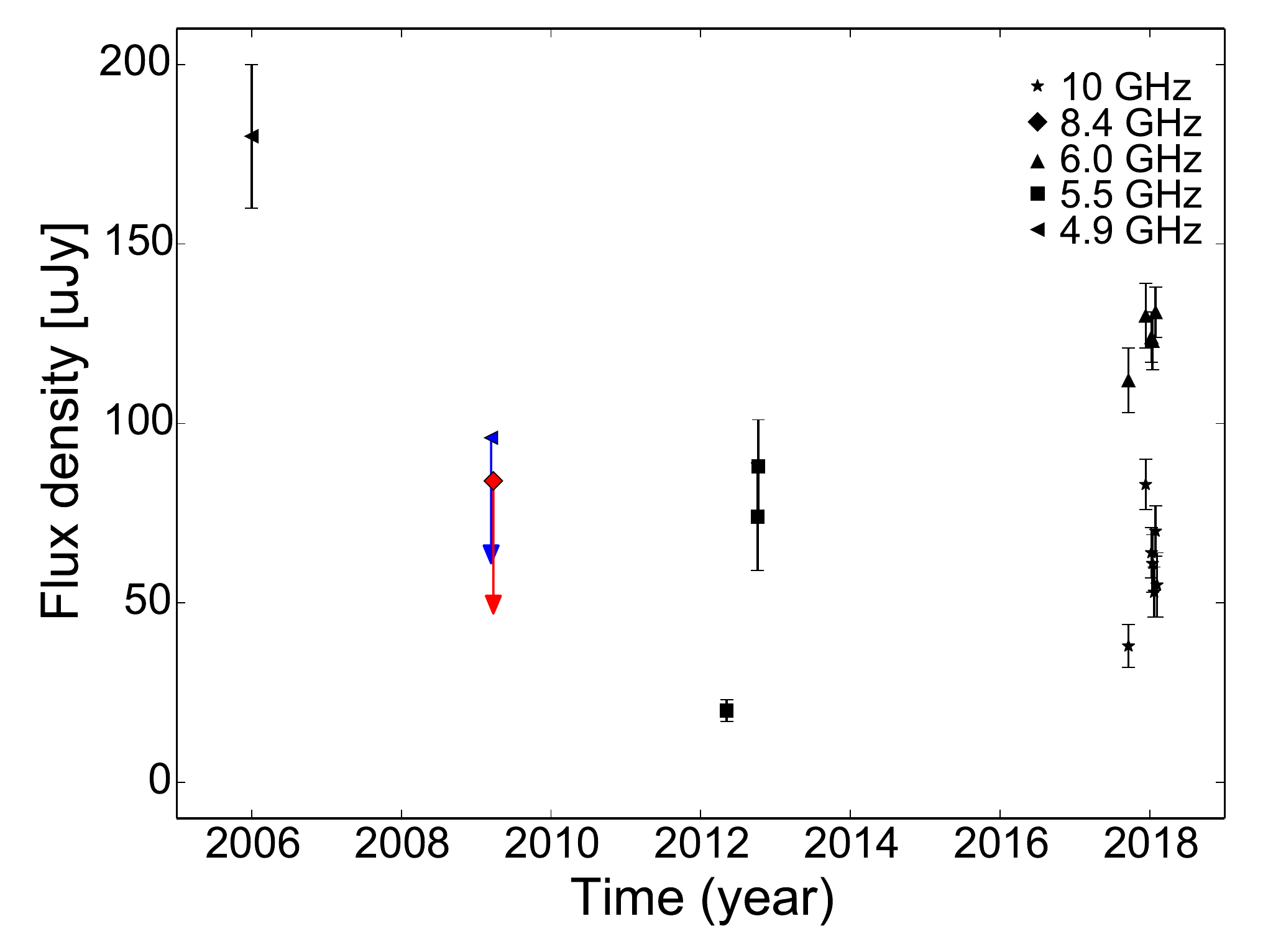}
   \caption{VLA J181335.1-174957 flux densities, in radio continuum, at the different observed frequencies as a function of time. 
   Values of flux densities before 2010 from \cite{dzib2010}. Upper-limits in 2009 observations are in colors just to distinguish the
   arrows.}
   \label{fig:FT}
\end{figure}

VLA J181335.1$-$174957 was detected in all the VLA observations presented in Sect. 2,
confirming the existence of the variable radio source reported by \cite{dzib2010}.
In all the observations, the radio continuum source is compact and point like  
(see Figure~\ref{fig:SNR}). The  measured flux densities and the image properties 
of each epoch are presented in Table~\ref{tab:vlapsr} and Figure~\ref{fig:FT}. Other 
sources detected in the field of these new VLA observations are presented and briefly 
discussed in Appendix~\ref{Apen:RS}. We note, in particular, that the angular resolution
and sensitivity of the observation obtained on 2012 May 06 enabled the detection of
the diffuse emission of SNR G12.82+0.02 (also see Figure~\ref{fig:SNR}). 

The images in the Stokes V parameter imply that VLA J181335.1$-$174957 has a 
right circular polarization of $\pi_c\sim$50\% (column 10 in Table~\ref{tab:vlapsr}).
In those epochs where C and X bands were observed, we also calculated the spectral 
indices ($\alpha$; $S_\nu\propto\nu^\alpha$) between these bands; they are shown in column 9 of Table~\ref{tab:vlapsr}. 
The source has a negative spectral index ($\langle\alpha\rangle=-1.3\pm0.1$), high circular polarization and it is 
variable. All this clearly indicates that the radio emission of the source is non-thermal.

The 24-minute Effelsberg 100-m telescope observations were searched for radio pulsations in two ways. We  folded the baseband data with the published $P$ and $\dot {P}$ \citep{halpern2012} and searched this data to a DM of 3500. Young pulsars are known to undergo sudden spin up in the rotation rate known as glitches, and this peaks for neutron stars with characteristic age, $\tau_c \approx 10$~kyr \citep{elsk2011}. To rule out spin period changes owing to glitches,  we carried out a full periodicity search with PRESTO, sensitive to a DM of 3496. Both these searches did not yield any pulsed radio emission
with 10$\sigma$ upper limits of $65\,\mu$Jy for a pulsar duty cycle of 10\%.  These levels are 
comparable with the non-detections with the GBT by \citet{halpern2012}.

\section{Discussion}
  
The flux densities of VLA~J181335.1$-$174957, presented in Table~\ref{tab:vlapsr} and 
Figure~\ref{fig:FT}, show that it was active from 2017 September to  2018 February 4 
with no strong variations. The variability coefficients\footnote{The ratio of the 
standard deviation to the mean.} for C- and X-band during these epochs were 0.06 and 0.22, 
respectively, indicating only moderate variability. One of our new Effelsberg observations 
to search for pulsed emission occurs only a week later after our last VLA observation. 
From the mean emission flux at 10.0~GHz of $\langle S_{10GHz}\rangle=57\pm3\,\mu$Jy and the spectral index 
of $\langle\alpha\rangle=-1.3\pm0.1$, the predicted flux density at 1.36~GHz of 
VLA~J181335.1$-$174957 is $762\pm14\,\mu$Jy, this is much higher than our 10$\sigma$ upper limits
of 69$\,\mu$Jy. 
As the continuum source showed only moderate variability the previous five months, its 
detection with the VLA and non-detection with Effelsberg support the idea that the continuum 
radio source is not the time-averaged pulsed emission from the pulsar, as suggested 
by \cite{halpern2012}. This also suggests that the radio  emission beam of the pulses do 
not point in the sightline toward the Earth.  We can not, however, discard high levels of 
scattering that will mainly affect low frequencies, leaving the pulsed emission undetected in the 
1.4--2.0 GHz observations (Camilo et al. in preparation).
The question still open is: what produces 
the observed continuum compact radio source VLA~J181335.1$-$174957?  We will now discuss some other possibilities to the origin of the radio continuum compact source.

A first hypothesis is that the radio emission comes from the pulsar wind (PW).
It is thought that most of the spin-down luminosity of pulsars such as PSR J1813-1749
is dissipated via a magnetized wind populated by relativistic electrons and positrons
\citep{rees1974}. At distances between 0.1 and 1~pc from the pulsar, this wind produces an extended 
PWN \citep{gaensler2006}. An upper-limit to the size of VLA~J181335.1$-$174957
of $\sim0\rlap{.}''5$  can be estimated from the highest angular resolution data presented 
earlier. At a distance of 4.8 kpc \citep{halpern2012} 
this corresponds to 2400 AU (0.01 pc) and is an order of magnitude more compact than the smallest known PWNe.
This means that the compact radio emission source may trace the inner parts of the PW.
The PWNe are detected across the electromagnetic spectrum
in synchrotron and inverse Compton emission and in optical emission lines when the pulsar
wind shocks the surrounding medium \citep{gaensler2006}. In contrast, there is scarce 
observational or theoretical information and predictions about the PW itself. Even for the 
Crab pulsar, little or no synchrotron emission is seen from the wind in the volume 
immediately surrounding the pulsar \citep{schmidt1979}, indicating that the wind itself
is not luminous. In general, it is believed that this ``cold'' PW cannot be 
observed before its termination in the PWN \citep{buhler2014}. \cite{istomin2011} 
argues that the PW could produce detectable synchrotron emission in transient pulsars and \cite{jones1977} 
show that significant amounts of circular polarization can be produced in homogeneous 
synchrotron sources. However, as there is no theoretic background to predict the 
flux densities and the properties of the radio emission of the PW in all pulsars, 
is not possible now to test our results on VLA J181335.1$-$174957. Theoretical studies of 
radio emission from PW will be highly important for this and future similar works.

Pulsars can also produce radio emission in their off-pulse states, additionally to the 
pulse emission \citep{basu2011}. There is evidence that this off-pulse radio emission has 
a magnetospheric nature and, when present, is typically 10 times weaker than the pulsed 
radio emission \citep{basu2012}. A possible interpretation of the steady radio emission 
detected here is that it corresponds to such off-pulse radio emission, but that in the 
present case of PSR J1813--1749, the pulse emission itself is not detected because (as 
mentioned earlier) the pulse emission beam does not point toward the Earth. The measured 
spectral index for VLA J181335.1$-$174957 ($\langle\alpha\rangle=-1.3\pm0.1$) is consistent 
with the off-pulse emission to other pulsars \citep[$\langle\alpha_{OFF}\rangle=-1.4$;][]{basu2012}
as is its variability. The off-pulse emission has only been investigated in old long period 
pulsars, which is not the case of PSR J1813-1749. The relation of this mechanism with source 
VLA J181335.1$-$174957 is still open.

A third possible interpretation of the radio emission detected here is  that the pulsar 
has a young, low-mass stellar companion that is producing gyrosynchrotron emission. 
The star that produced the pulsar was formed $\leq10$ million years ago 
and it is possible that a low-mass star may be associated with the pulsar.  Low-mass stars are 
frequently magnetically active and may produce significant amounts of gyrosynchrotron emission
\citep[e.g.,][]{dzib2013}.  This emission is typically strongly variable,
on scales of hours to days, with a wide range of spectral indices, usually in the range from 
$-2$ to $+2$, and 
they may exhibit significant amounts of circular polarization. There is a precedent to this 
possibility.  In the case of {\it PSR} J1603-7202,
it has been shown that its low-mass companion produces significant amounts of circularly polarized
radio continuum emission \citep{manchester2004}. However, in this case the large age of the pulsar,
$\geq$10 Gyr \citep{lorimer1996}, implies that the low mass companion may be a white dwarf.
Even though the young low-mass companion hypothesis provides a plausible explanation, there are 
also some inconsistencies. Radio emission from magnetically active 
stars is usually strongly variable in flux, in circular polarization, and in spectral index. This
is inconsistent with the period of moderate variability of VLA~J181335.1$-$174957 documented here. 
It should be pointed out, however, that there is a set of young intermediate mass stars that show
circular polarization and no detectable variability \citep[e.g., the sources GBS-VLA J162634.17-242328.4 
and GBS-VLA J162749.85-242540.5 in][]{dzib2013}.
On the other hand, the orbit between the pulsar and the putative stellar companion would have to 
be nearly in the plane of the sky since no periodicities were reported in the X-ray
timing observation of \cite{halpern2012}. Furthermore, at distances  $\geq4.8$~kpc \citep{halpern2012}
this would be one of the brightest gyrosynchrotron stars known, comparable only to 
the brightest similar sources in Orion \citep{zapata2004}. 

Finally, the solution to the enigma may be surprisingly simple. Camilo et al. (in preparation) have proposed that PSR J1813-1749 is the most scattered pulsar known and that the pulses become evident only at high frequencies.

\section{Conclusions}

We presented new deep interferometric VLA observations and single dish Effelsberg observations 
of the continuum radio source VLA J181335.1$-$174957, which has been related to the high energy 
pulsar PSR J1813-1749 and to the TeV source HESS J1813-178. The radio continuum source was detected
in 10 different epochs with the VLA. The observations showed that 
VLA J181335.1$-$174957 is variable, with a circular polarization of $\sim50\%$ and a spectral index 
of $\langle\alpha\rangle=-1.3\pm0.1$. These parameters indicate that the radio emission has a non-thermal
nature. The observations with the Effelsberg telescope, however, find no pulse emission and support the 
idea suggested by \cite{halpern2012} that the steady VLA radio source is not the integrated radio pulse emission. 

We have discussed three different possibilities on the nature of VLA J181335.1$-$174957. 
The most promising explanations of the radio continuum emission are that it corresponds 
either to the pulsar wind, or to magnetospheric emission from the pulsar (also known as 
off-pulse emission). The former explanation is very exciting since it would indicate the first detection 
of a pulsar wind.  However, the information of pulsar winds and the off-pulse emission
is still scarce, so we cannot currently favor or discard either of these possibilities. 
A third explanation is that PSR J1813-1749 has a low-mass stellar companion which  
produces the non-thermal radio emission. In this case, VLA J181335.1$-$174957 is not directly 
related to the pulsar. However, there is no independent evidence (e.g.\ based on timing) 
that PSR J1813--1749 has a companion. The enigma may have a surprisingly simple solution: 
Camilo et al. (in preparation) have found that PSR J1813-1749 is the most scattered pulsar 
known and that observations made at higher frequencies are consistent with the radio emission 
being the time-averaged pulsed emission.

\acknowledgements 
We thank the referee Jules Halpern for his suggestions that improve the 
readability of the manuscript, and for sharing his results prior publication. 
L.R. and L.L. acknowledges the financial support of DGAPA, UNAM (project IN112417), and CONACyT, M\'exico. 
S.-N.X.M. acknowledges IMPRS for a Ph.D. research scholarship.
The National Radio Astronomy Observatory is a facility of the National Science Foundation operated under 
cooperative agreement by Associated Universities, Inc. This paper was partly based on observations with 
the 100-m telescope of the MPIfR (Max-Planck-Institut f\"ur Radioastronomie) at Effelsberg. {We thank Alex Kraus for scheduling these observations at the Effelbserg 100m in short notice.}

\facilities{VLA,  Effelsberg Telescope} 
\software{ CASA \citep{mcmullin2007}, BLOBCAT \citep{hales2012}, PSRIX \citep{lkg+2016}, 
NE2001 \citep{ne2001}, and YMN16 \citep{ymn2016}.}

\appendix
\section{Other radio sources}\label{Apen:RS}

In order to detect the weakest sources in the field we combined the observation of the 
project 17B-028, to produce two deep images (one for each frequency band). 
The final noise level in both images is $\sim4\,\mu$Jy. Then, following 
\cite{medina2018}, we have used the BLOBCAT software \citep{hales2012} to
search for peaks above five times the noise level. To consider a source as real we 
use the following criteria: (i) the source has a counterpart at any other wavelength,
or (ii) its signal to noise ratio is above seven.

Under the above criteria, we detected 27 radio sources from which seven are 
massive stars, and another is related to a sub-mm source. The remaining sources 
do not have any reported counterpart. Most of the detected radio sources are compact or slightly 
resolved but two (VLA J181328.00-174958.0 and VLA~J181329.58-174829.4) are clearly extended. 
All detected radio sources and their measured fluxes are listed in Table~\ref{tab:RS}.
The spectral index was calculated for compact sources detected in both bands, and it
is also shown in Table~\ref{tab:RS}.
Now, we will briefly discuss the most interesting sources.

\begin{figure*}[!bth]
   \centering
   \begin{tabular}{cc}
  \includegraphics[height=0.5\textwidth, trim=20 330 105 40, clip]{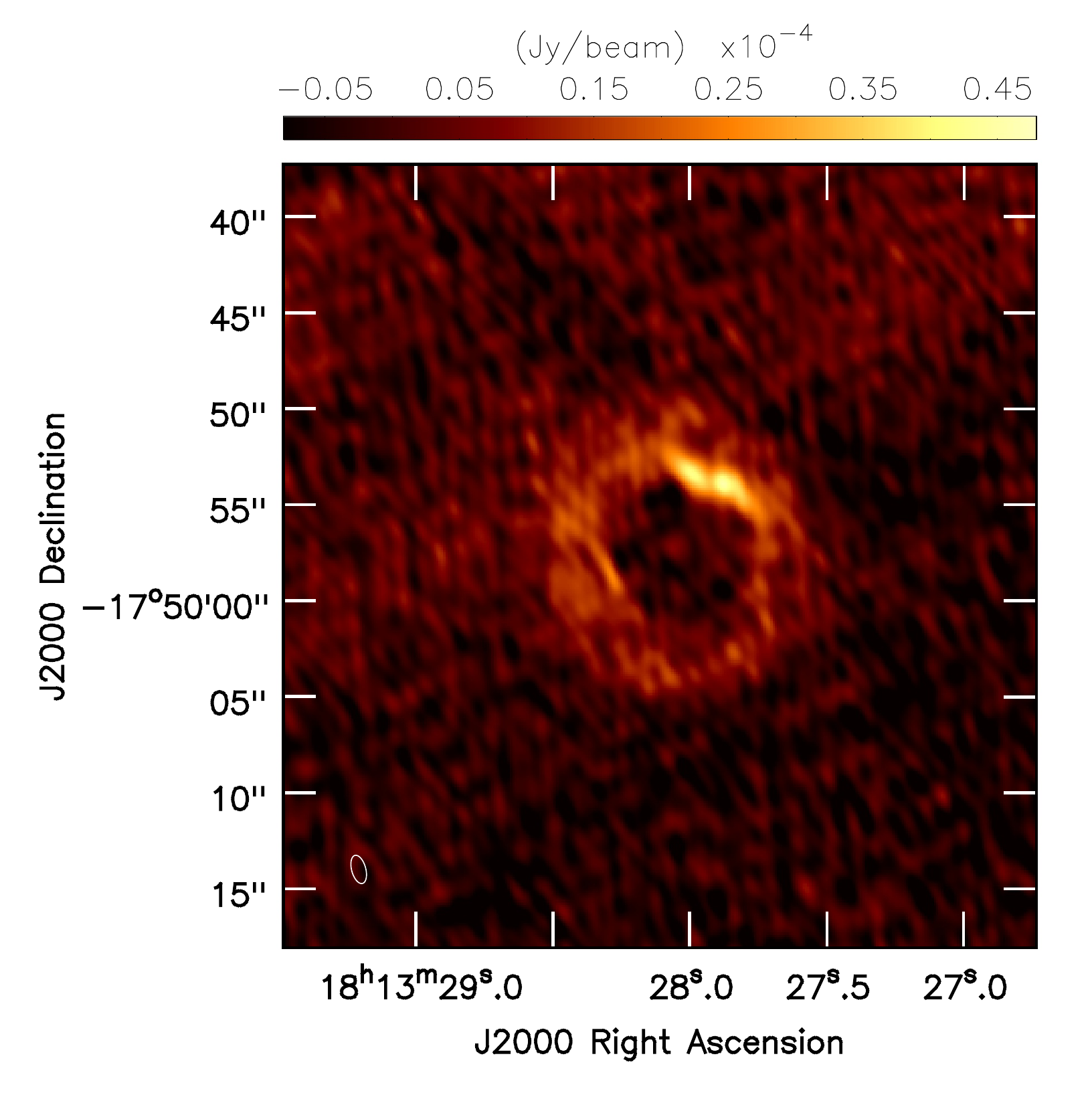}&
  \includegraphics[height=0.5\textwidth, trim=35 330 90 40, clip]{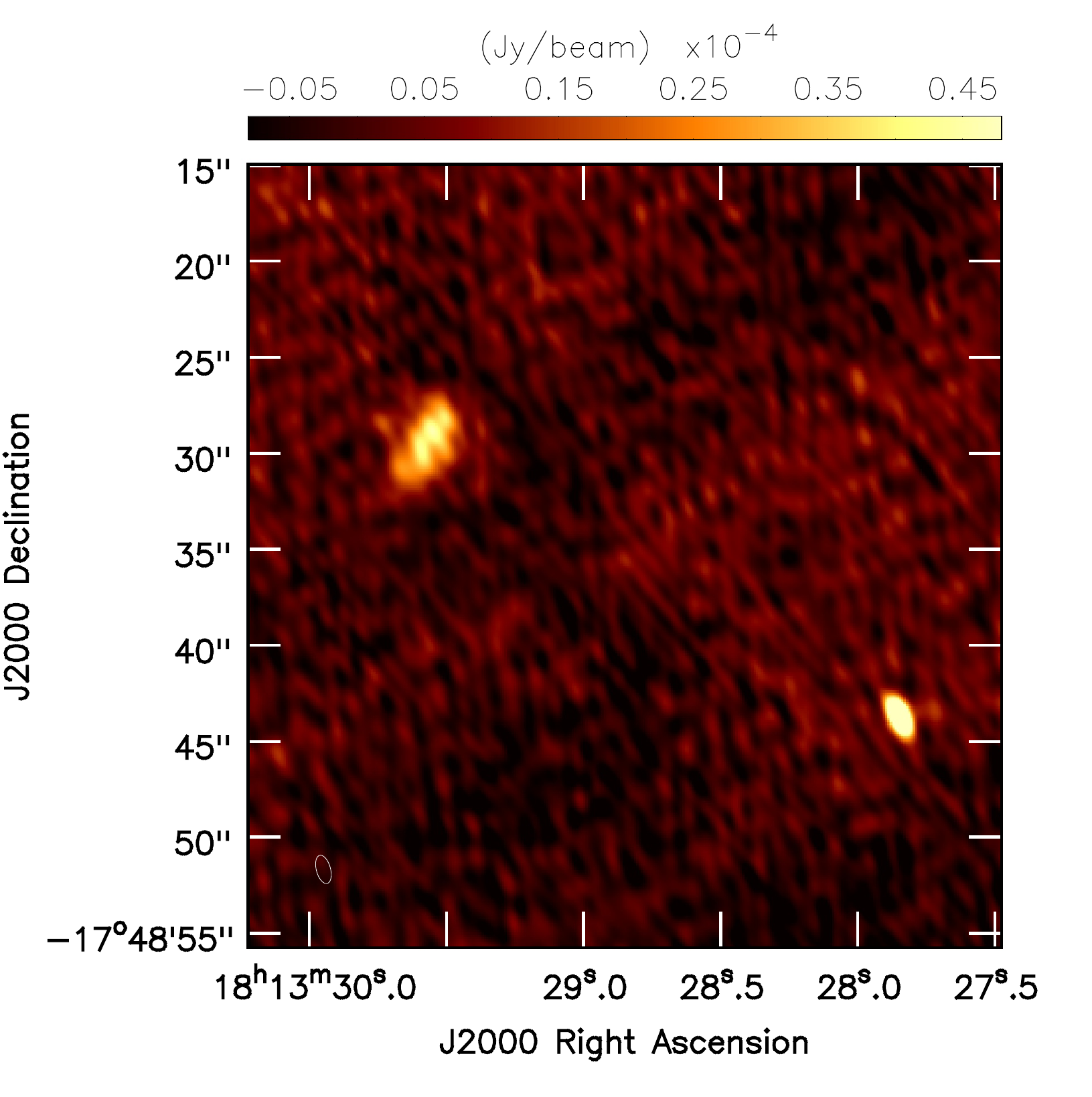} \\
  \end{tabular}
   \caption{{\it Left:} C-band radio image of the shell like radio source VLA~J181328.00-174958.0.
   {\it Right:} C-band radio image of the VLA~J181329.58-174829.4 and VLA~J181327.85-174843.7 region.}
   \label{fig:wcr}
\end{figure*}

{\it VLA~J181328.00-174958.0} appears as an irregularly shaped source in the 
X-band image. In the C-band image, on the other hand, a clear shell like morphology 
is recovered (see Figure~\ref{fig:wcr}, left panel). This source is surrounded by evolved massive 
stars, so its location, and morphology suggest that it may be a SNR. We
cannot discard the possibility, however, that it is an HII region, a planetary nebula, or the nebula around 
an evolved massive star \citep[i.e.,][]{duncan2002}. The smaller flux density at X-band, 
as compared to C-band, supports a supernova remnant nature, but the modest short spacing 
coverage of our data could introduce systematic errors in the flux densities of the 
extended sources. More information is needed to clarify its nature.

{\it VLA~J181348.28-174539.2} is associated with the sub-mm source 
AGAL~G012.904-00.031 \citep{contreras2013} and with the methanol maser
MMB~G012.904-00.031  \citep{green2010}. Association between dense sub-mm and 
methanol masers are usually related to massive star formation \citep{james2013},
suggesting that VLA~J181348.28-174539.2 is related to a massive young stellar 
object.

{\it VLA~J181329.58-174829.4} is a resolved radio source with an arc-like
morphology (see Figure~\ref{fig:wcr}, right panel). Wind collision regions 
(WCR) in massive binary stars may produce radio
emission with this kind of structure \cite[e.g.,][]{ortiz2011}. Interestingly, the apex of the arc 
points to another radio source  {\it VLA~J181327.85-174843.7}. Thus we suggest that 
the first source is a WCR and the second source is a massive star. Its massive
companion should be closer and at the east side of VLA~J181329.58-174829.4.

The {\it massive stars} with detected radio counterparts in Table~\ref{tab:RS},
are part of a young massive cluster discovered by \cite{messineo2008}. To our 
knowledge this is the first detection at radio frequencies of these massive stars.

\begin{table*}[b]
\small
\centering
\renewcommand{\arraystretch}{0.9}
\caption{Properties of the other radio sources detected in the VLA observations.}
\begin{tabular}{lcccccccc}\hline\hline
       &      & Spectral     &    $S_{\nu,{\rm C}}$             &  $S_{\nu,{\rm X}}$   &           \\
VLA Name& Other name & type    &   ($\mu$Jy)  & ($\mu$Jy) &$\alpha$\\
\hline
J181310.95-175141.9 & ...                     & ... & $1141\pm71$ &  ... & ... \\
J181312.88-174908.5 & ...                     & ... & $176\pm22$  & ... & ... \\
J181314.22-175343.3 & 2MASS~J18131420-1753434 & WN7b & $1067\pm76$ &  ... & ... \\
J181318.12-175327.1 & ...                     & ... & $810\pm45$ &  ... & ... \\
J181319.08-175258.4 & 2MASS~J18131908-1752585 &O8-O9If & $423\pm25$ &  ... & ... \\
J181319.48-174813.3 & ...                     & ... & $1075\pm54$ &  549$\pm$88 & $-1.3\pm0.3$\\
J181321.00-174947.1 & 2MASS~J18132098-1749469 & LBV & $703\pm36$ &  1042$\pm$55 & $0.8\pm0.1$\\
J181321.24-175141.4 & ...                     & ... & $134\pm10$ &  ... & ... \\
J181322.51-175350.3 & 2MASS~J18132248-1753503 & WN7 & $110\pm14$ &  ... & ... \\
J181322.54-174504.8 & ...                     & ... & $917\pm53$ &  ... & ... \\
J181324.44-175256.9 & 2MASS~J18132443-1752567 & B0-B3 & $119\pm10$ &  ... & ... \\
J181326.48-175249.5 & 2MASS~J18132647-1752495 & B0-B3 & $36\pm7$ &  $<70\pm23$ & ... \\
J181327.85-174843.7 & ...                     & ... & $174\pm10$ &  140$\pm$8 & $-0.4\pm0.2$\\
J181328.00-174958.0 & ...                     & ... & $865\pm43$ &  $190\pm10$ & ... \\
J181328.73-175142.0 & ...                     & ... & $44\pm5$ &  39$\pm$5 & $-0.2\pm0.3$\\
J181329.58-174829.4 & ...                     & ... & $369\pm19$ &  $167\pm16$ & ... \\
J181332.15-174728.0 & ...                     & ... & $97\pm7$ &  49$\pm$7 & $-1.3\pm0.3$\\
J181333.82-175148.5 & 2MASS~J18133381-1751485 & O6-O7f & $23\pm4$ &  17$\pm$4 & $-0.6\pm0.6$\\
J181335.08-175027.1 & ...                     & ... & $27\pm4$ &  ... & ... \\
J181339.12-175037.2 & ...                     & ... & $22\pm4$ &  ... & ... \\
J181340.11-175143.8 & ...                     & ... & $38\pm5$ &  24$\pm$5 & $-0.9\pm0.5$\\
J181341.20-175038.6 & ...                     & ... & $29\pm4$ &  ... & ... \\
J181344.73-174759.2 & ...                     & ... & $320\pm17$ &  191$\pm$14 & $-1.0\pm0.2$\\
J181346.53-174856.4 & ...                     & ... & $97\pm8$ &  ... & ... \\
J181347.23-175118.3 & ...                     & ... & $37\pm6$ &  63$\pm$14 & $1.0\pm0.5$\\
J181348.28-174539.2 & AGAL~G012.904-00.031 & ... & $234\pm22$ & ...  & ... \\
J181350.60-174807.8 & ...                     & ... & $1191\pm60$ &  771$\pm$89 & $-0.9\pm0.2$\\
\hline\hline
\label{tab:RS}
\end{tabular}
\end{table*}

\bibliographystyle{aa}
\bibliography{references}

\end{document}